# Comparative Study of Simulated Nebulized and Spray Particle Deposition in Chronic Rhinosinusitis Patients

Zainab Farzal, MD[1], Saikat Basu, PhD,[1] Alyssa Burke,[2] Olulade O. Fasanmade, BS,[1] Erin M. Mamuyac, MD,[1] William D. Bennett, PhD,[2] Charles S. Ebert Jr, MD, MPH,[1] Adam M. Zanation, MD,[1] Brent A. Senior, MD,[1] Julia S. Kimbell, PhD[1]

[1]Department of Otolaryngology/Head and Neck Surgery, University of North Carolina at Chapel Hill

[2]Center for Environmental Medicine, Asthma, and Lung Biology, University of North Carolina at Chapel Hill

**Corresponding Author:**

Zainab Farzal, MD

170 Manning Drive, CB #7070

Physician's Office Building Room G-190

Chapel Hill, NC, 27599

Phone: 919-966-3342

Fax: 919-933-7941

Email: ZFarzal1@gmail.com

Financial Conflict of Interest Disclosure: None

Presented as an oral presentation at the Annual American Rhinologic Society (ARS) Meeting in Atlanta, Georgia on October 5-6, 2018.

Short Title: Nebulizers Computational Analysis

# Comparative Study of Simulated Nebulized and Spray Particle Deposition in Chronic Rhinosinusitis Patients

**Introduction:** Topical intranasal drugs are widely prescribed for Chronic Rhinosinusitis (CRS), although delivery can vary with device type and droplet size. The study objective was to compare nebulized and sprayed droplet deposition in the paranasal sinuses and ostiomeatal complex (OMC) across multiple droplet sizes in CRS patients using computational fluid dynamics (CFD).

**Methods:** Three-dimensional models of sinonasal cavities were constructed from computed tomography (CT) scans of three subjects with CRS refractory to medical therapy using imaging software. Assuming steady-state inspiratory airflow at resting rate, CFD was used to simulate 1-120 μm sprayed droplet deposition in the left and right sinuses and OMC with spray nozzle positioning as in current nasal spray use instructions. Zero-velocity nebulization simulations were performed for 1-30 μm droplet sizes, maximal sinus and OMC deposition fractions (MSDF) were obtained, and sizes that achieved at least 50% of MSDF were identified. Nebulized MSDF was compared to sprayed droplet deposition. We also validated CFD framework through *in vitro* experiments.

**Results:** Among nebulized droplet sizes, 11-14 μm droplets achieved at least 50% of MSDF in all six sinonasal cavities. Five of six sinonasal cavities had greater sinus and OMC deposition with nebulized droplets than with sprayed droplets at optimal sizes.

**Conclusions:** Nebulized droplets may target the sinuses and OMC more effectively than sprayed particles at sizes achieving best deposition. Further studies are needed to confirm our preliminary findings. Several commercial nasal nebulizers have average particle sizes outside the optimal nebulized droplet size range found here, suggesting potential for product enhancement.

**INTRODUCTION**

Chronic rhinosinusitis (CRS) is a pervasive condition with a well-characterized, extensive public health burden.[1] Symptoms include nasal obstruction, facial pressure, nasal discharge, and loss of smell. Additionally, CRS is associated with worsened quality of life, decreased productivity, and results in over $9 billion yearly in healthcare costs and more than $13 billion in societal costs in the United States.[2-4] Functional endoscopic sinus surgery (FESS) is typically undertaken after medical therapy fails. Initial medical management includes antibiotics, nasal saline irrigations, and topical nasal or oral steroids.

Topical treatment allows for direct drug delivery and reduction of potential systemic side effects, particularly for long-term use.[5,6] Most patients are prescribed intranasal steroids in a spray formulation as part of their medical regimen. However, many patients with CRS eventually undergo surgery owing to a disease burden unresponsive to medical treatment. Intranasal spray delivery is based on many patient and drug-specific factors, and may be minimally effective if these factors are not optimized. Patient-specific factors include an understanding of head positioning, spray nozzle direction or positioning within the nasal cavity, and inhalation at time of spray. Drug-specific factors include drug particle size distribution, spray emission speed, and shot weight. Enhancing topical drug delivery to target diseased areas may result in better treatment outcomes and potentially reduce the need for surgery.

Nebulizers may minimize several patient and drug-specific impediments to delivery that are encountered with nasal sprays, improving delivery to target sites. Since particles are aerosolized, nebulization permits widespread delivery throughout the nasal cavity including sinus ostia independent of device direction or angle encountered with nasal sprays. Additionally, most nebulizers use small particles which are likely to travel further to target sites instead of depositing in the anterior nose.[7-10] Several types of nebulizers are commercially available, including active and passive flow nebulizers. Active flow devices include nebulizers with particle release at a given velocity, while passive nebulizers rely only on the negative pressure generated by inhalation.

Although nebulizers may hold great potential, possibly reducing the need for surgery and associated morbidity or costs, little evidence exists to support their inclusion in the main CRS medical armamentarium. To date, primarily radiolabeled tracers or gamma scintigraphy methods have been used to assess nebulizer drug delivery to the sinuses.[11-14] Only two prior studies have compared spray and nebulizer particle deposition in the sinonasal cavity demonstrating greater sinus deposition of nebulized particles.[11,12] Computational fluid dynamics (CFD) methods enable simulation of airflow and particle delivery to the sinonasal cavity, and have been validated with physical models.[15-19] Although CFD is emerging as the standard for assessing airflow and particle deposition, only one prior study has used CFD to analyze sinonasal nebulized particle deposition.[17] The objective of this study was to compare nebulized and sprayed droplet deposition in the paranasal sinuses and ostiomeatal complex (OMC) across multiple droplet sizes in patients with CRS without nasal polyps using computational fluid dynamics (CFD). We hypothesize that sinus and OMC particle deposition will be higher with nebulizers compared to sprays.

**METHODS**

**Patient Selection**

The study was approved by the University of North Carolina Institutional Review Board (IRB #10-0556). Pre-operative CT scans belonging to three individuals with CRS without nasal polyposis were used for the three-dimensional sinonasal reconstructions. Three patients (2 males, 1 female) who were surgical candidates after failed medical management were specifically chosen to assess drug delivery with nebulizers compared to topical sprays in a diseased state with unaltered anatomy (**Table 1**). At the same visit, minute volumes were obtained during resting breathing using a portable respiratory inductive plethysmograph (LifeShirt®, VivoMetrics, San Diego, CA) and used for simulations.

**Three-Dimensional Model Development for CFD**

The de-identified CT scans were imported into Mimics™ 18.0 (Materialize, Inc., Plymouth, MI, USA) imaging software. The airway was reconstructed using an imaging radiodensity threshold range of -1024 to -300 Houndsfield units with hand editing to ensure correct anatomy. The models were approved by multiple otolaryngologists for precision. Subsequently, the models were exported into the computer-aided design and meshing software ICEM-CFD™ 15.0 (ANSYS, Inc., Canonsburg, PA, USA). Boundary surfaces were created including nostrils and an outlet below the nasopharynx. The model was divided into anatomic regions including the anterior nose, main nasal cavity, middle turbinate, sinuses (maxillary, frontal, ethmoids, and sphenoid), ostiomeatal complex (OMC), and nasopharynx (**Figure 1A-C**). Based on previously validated methods, computational meshes of approximately 4 million tetrahedral cells with three 0.1-mm thick prism layers along airway walls were created (**Figure 1D**).[20,21] Mesh quality was verified by ensuring that if present, the number of distorted low-quality elements was less than 40 (0.001%) .

**Airflow Simulations**

Steady-state, laminar, inspiratory airflow simulations were performed using Fluent™ (ANSYS, Inc., Canonsburg, PA, USA), at twice the resting breathing minute volume for each subject as measured at time of recruitment (**Table 1**). The numerical methods followed previously published work in the literature.[17,22,23]

**Spray and Nebulized Particle Deposition CFD Simulations**

Nasal sprays were simulated as series of solid-cone injections in Fluent™, one for each droplet diameter from 1 to 120 μm in 1μm increments, using a cone angle measured 3 cm above the sprayer (63.3°) and velocity based on patient-specific breathing rates. Injections were released from a point positioned 5 mm vertically into the nose from the nostril centroid with the head tilted forward at 22.5°. Deposition fractions were computed for each region in the models for each droplet size, using 5000 streams per injection.

Nebulizers were simulated using the "surface" injection type in Fluent™. Ten thousand nostril surface release points were created. The authors were concerned that particles released too close to nostril edges *in vivo* may adhere to the nasal sill, columella, or anterior most aspect of the septum. To exclude these particles in the simulations so that deposition patterns in the computational study are not artificially inflated, a scale factor of 0.95 was used to circumferentially reduce the nostril surface used for particle release. Due to exclusion of the 5% circumferential surface area, a range of 8868 to 9868 particles (from the original 10,000 spanning the original nostril surface) were released in the simulations. Nebulized droplet particles were simulated at zero velocity for monodisperse particle sizes ranging from 1-30 μm in the left and right sinuses and OMC. The simulations were performed in an upright head position.

**Validation Modeling**

An *in vitro* experiment was performed to validate nebulizer results. A 3-D printer was previously used to print the sinonasal airway and external nose for Subject 3. The nose was made from a pliable material with a similar texture to the external nose. The airway past the vestibule was made from a firmer material, representative of bony anatomy (**Figure 2**).

**Nebulized Aeresol Delivery for the *in vitro* Model**

The 3-D printed model was exposed to a nebulized solution of sodium pertechnetate-99 ($^{99m}$Tc) (10 millicuries in 0.4 mL of normal saline). We used a PARI LL jet nebulizer (Starnberg, Germany) modified by removing the lower third of the internal baffle to allow generation of aerosol particles (mass median aerodynamic diameter of 9.5 μm). Flow was drawn through the nebulizer and a single nostril (right side) of the model at a constant 6 L/min airflow in the inspiratory direction for approximately 100 seconds (**Figure 2**). Under these conditions we estimated 200 μL may be delivered to the model, providing sufficient radioactivity for gamma scintigraphy scans (described below) while minimizing displacement of fluid on the internal model surfaces prior to scanning.

**Gamma Scintigraphy Scans**

A gamma scintigraphy camera was used to record images of the models before and after administration of the nasal spray. Each scan, lasting three minutes, consisted of two images taken simultaneously, recording either Technetium or Americium energy levels. Three point sources of Americium were used as markers within the model to facilitate image processing. Five scans were taken for each experiment to examine deposition from the front and side of the model. Two background scans (front of the model and side of the model) were taken first. All "side" scans were taken on the right side of the model where the drug was administered. All front scans were performed without the soft nose attached. For side scans, the camera was rotated to 90 degrees and the side of the model was pressed flush to the camera. The head of the model was tilted forward at a 22.5 degree angle from upright.

After completion of the background scans, the nasal aerosol was administered into the indicated nostril as described above. Immediately following aerosol delivery, a scan was taken from the side, with the soft "anterior nose" still attached. The soft nose was then removed, and activity was measured. Another scan from the side was taken with the soft nose removed. The final scan was a front scan taken in the same manner as described for the background scan.

**Gamma Scintigraphy Image Processing**

Processing of the scans was completed in ImageJ (National Institutes of Health). A region of interest (ROI) was constructed for each model, and for each view (front or side). Construction of each ROI was designed relative to the locations of the Americium markers. The side view was divided into vertical bins (columns/coronal planes) and horizontal bins (rows/transverse planes), as well as a bin for the filter. Data on the number of counts in each ROI was imported into Microsoft Excel 2017 and the difference between the scans after spray and the background scans was used to determine the amount of activity in each region due to the labeled nasal aerosol.

**CFD Simulation for Gamma Scintigraphy Comparison**

For CFD simulation, a round inlet was created in subject 3's right nostril and used as the release surface for nebulized droplets. Nine micron sized particles were simulated, consistent with the nebulizer particle size used in the *in vitro* validation modeling. A total of 18,000 particles were released from the surface for visualization and comparison to the *in vitro* experiment. All other simulations conditions were identical to the 1-30 micron particle size nebulizer simulations.

**Visualization and Statistical Analysis**

Visualization and analysis of results were performed in the post-processing software package FieldView™ 16 (Intelligent Light, Lyndhurst, PA). Sinus and OMC deposition fractions were defined as the number of total particles depositing in the sinuses and OMC divided by the total particles released. The particle size achieving maximal sinus and OMC deposition fractions (MSDF) was determined for both sprays and nebulizers. Sizes that achieved at least 50% of MSDF were also identified (**Figure 3**). Nebulized MSDF was compared to sprayed droplet deposition for the same size, and vice versa. Additionally, nebulizer and spray post-nasal penetration fraction was determined reflecting the fraction of droplets that traveled and deposited past the anterior nasal cavity. Total nebulizer and spray deposition fractions were also calculated for each particle size representing the fraction of particles that deposited in the sinonasal cavity (ie. did not escape through the nasopharynx). To statistically check the physical basis of the nebulized numerical predictions, we ran Spearman's rank correlation test between the particle deposition fractions in the CFD simulations and the *in vitro* experiments. The test compared the deposition trends in the coronal, sagittal, and axial compartments of a representative model (Subject 3 RHS), and the reported numbers included Spearman's rho and two-tailed *p values*.

**RESULTS**

Among nebulized particles, the droplet sizes achieving maximum sinus and OMC deposition (MSDF) among the six nasal sides ranged from 11 to 21 microns (**Table 2**). Particles within the range of 11-14 microns consistently achieved at least 50% of MSDF in all six sinonasal cavities (**Table 2, Figure 4**). The percent of nebulized droplets depositing past the anterior nasal cavity (defined as penetration) ranged from 55.6% to 96.7% at particle sizes achieving MSDF (**Table 2**).

MSDF occurred at spray droplet sizes ranging from 4 to 13 microns among the six nasal sides (**Table 2**). No common particle size range was noted for particle sizes reaching at least 50% MSDF. Five of 6 nasal sides had overlap in the 10 to 11 micron particle range for 50% MSDF. The spray droplets percent penetration past the anterior nasal cavity ranged from 20.7% to 99.8% at particle sizes achieving MSDF (**Table 2**).

When comparing sprayed and nebulized droplets in each sinonasal cavity, nebulizers droplets had greater MSDF compared to sprays in all but one case (subject 2 LHS, 23.1% nebulized vs. 35.3% sprayed, **Table 2, Figure 5**). The nebulized MSDF achieved in four of six cases was greater than the sinus and OMC deposition fraction or percentage of the same size spray particle; in one case, deposition was nearly identical, and the spray had greater deposition in the other case (**Table 2**).

The distribution fraction for all particle sizes to the OMC region and maxillary, ethmoid, sphenoid, and frontal sinuses is represented in **Figure 6** for each subject. The OMC region had the highest spray and nebulizer particle deposition in most cases across subjects with the exception of subject 3 who had high nebulized particle deposition to the ethmoid sinuses in addition to the OMC (**Figure 6**).

Nebulizer and spray post-nasal penetration and total deposition fractions are shown in Figure 7. Larger sizes had lower penetration, depositing in the nasal cavity. This trend was more evident in sprays which had a larger particle size distribution (**Figure 7**). The plots of total deposition fractions and particle size were sigmoidal. Five micron or smaller nebulizer and spray particles had small total deposition fractions (0 to 0.2) and escaped through the nasopharynx; almost all larger particles deposited reaching a deposition fraction of 1 for the largest sizes (**Figure 7**).

A visual comparison of the *in silico* (CFD) and *in vitro* (gamma scintigraphy) nebulized particle deposition for the purpose of CFD simulation validation is shown in **Figure 8** for a representative model (subject 3, RHS). Similarly high deposition signals are visible at the anterior nose (**Figure 8A**) and the middle nasal cavity region in the vicinity of the OMC and at and below the level of the ethmoid sinuses (**Figure 8B, 8C**). The particle deposition fractions for CFD and gamma scintigraphy models are shown in **Figure 9**, representing significantly similar distribution across regions. For particle deposits in the coronal segments, the Spearman's rank correlation test returned R = 0.83891 and the *p value* was 0.002 (**Figure 9A, 9B**). Comparison of the deposits in the sagittal segments resulted in R = 0.91539, with *p value* <0.001 (**Figure 9C, 9D**). Finally, for particle deposits in the axial segments, the test gave R = 0.90696 and the *p value* was <0.001 (**Figure 9E, 9F).** Therefore by established standards (with all the two-tailed *p values* << 0.05), the congruity between the CFD simulated particle transport and the physical experiments in the 3-D solid replicate could be considered statistically significant.

**DISCUSSION**

Our preliminary study highlights greater total OMC and sinus particle deposition from nebulizers compared to sprays in most cases. Additionally, there appeared to be a size 'preference' between sprays and nebulizers. Ideal particle size for nebulizers remained in a more consistent range (between 10-14 microns) across patients, but fluctuated with a wider range for sprays. Furthermore, nebulizer particles of the same size as the spray particle sizes achieving MSDF had greater OMC and sinus deposition. Our group has previously shown that nebulizers deliver drugs more posteriorly in the nasal cavity compared to sprays when other anatomic changes such as septal deviation are present.[24] However, this is the first study to use CFD to compare nebulized and spray particle delivery to the sinonasal cavity in CRS patients validated by gamma scintigraphy. Most prior studies assessing nebulizer deposition have used radiolabeled tracers and gamma scintigraphy as primary methods, which may yield less precise deposition data.[11-14]

Inhaled particles deposit in different regions of the upper and lower airway based on their size.[10,25] Our study confirmed prior research highlighting that particles that are 5 microns or less in size are more likely to deposit in the lungs, while those that are 10 microns or larger are more likely to deposit in the sinonasal cavity.[9,10,24,25] Interestingly, our results highlight a common nebulizer particle deposition range of 10-14 microns, congruent with prior studies.[24] On the other hand, optimal spray deposition ranges were highly variable. Furthermore, sprays included particles over 100 microns in size which tend to deposit anteriorly in the nasal vestibule due to their bulk and gravity, also verified in our study.

Although the particle sizes in commercial sprays are optimal for rhinitis, the condition for which most intranasal steroid sprays are FDA-approved, these sprays are less than ideal treatment for sinusitis. In fact, one might postulate that patients may not be receiving maximal medical treatment prior to surgical consideration because of the inherent suboptimal nature of topical drug delivery to diseased areas. Particle size distributions could be altered in new sprays specifically manufactured for CRS. However, as reflected in our study, the broader range of effective sizes across different patients would add an obstacle to determining new particle distributions for sprays. From this standpoint, nebulizers are a better vehicle of drug delivery due to a consistent particle size range at peak deposition to the OMC and sinuses. Additionally, nebulizers cover more surface area due to greater and more widespread aerosolization.[10,26] However, many commercial nebulizers include particles that are less than 10 microns or larger than the optimal range in our study (**Table 3**), suggesting potential for product enhancement.

Despite lack of evidence to date whether or not nebulizers are better treatment than sprays based on physical deposition characteristics, the decision not to use nebulizers in standard of care may be based on concern for affordability and patient compliance. Most nebulizers, with costs listed in **Table 3**, are more expensive than sinonasal sprays including Flonase® and Nasacort® which can be obtained over the counter. Additionally, sinonasal nebulizers may not be approved by many insurance companies. Nasal sprays are also more portable and easy to comply with, while nebulizers are bulkier (**Table 3**), and may require a plug-in set-up. Despite the inconveniences associated with nebulizers, if they are shown to be more effective treatment compared to sprays, they may enhance appropriate medical therapy, particularly benefiting patients with refractory symptoms.

This pilot study was not without limitations. Firstly, we simulated a passive nebulizer with particle release at the nostril surface with inhalation. Therefore, our simulations do not represent the sinonasal nebulizers that have baseline actuation velocities and variable particle size distributions that may alter delivery. Our simulations did not mimic some unique nose pieces used with some nebulizers and their insertion depth in the nares, which may also influence particle deposition. However, we hope that this pilot study will serve as a baseline comparison for future analyses with modifications accounting for the diverse characteristics of current nebulizers. Lastly, our work focuses on pre-surgical CRS patients who have failed medical management in an effort to improve topical drug delivery and avoid surgery. Additional studies are needed to assess nebulized drug delivery in patients following FESS.

**CONCLUSIONS**

In this pilot CFD study comparing particle deposition among single size nebulized and spray particles in patients who failed medical treatment, we determined that nebulized droplets may target the sinuses and OMC more effectively than sprayed particles at sizes achieving best deposition. The particle size range achieving the highest penetration of OMC and sinuses was consistent in the 10-14 microns range for nebulizers, but varied greatly for sprays. Additionally, this optimal nebulized droplet size range was different from those of several commercial nasal nebulizers, suggesting potential for product enhancement in the future.

## ACKNOWLEDGEMENTS

Research reported in this study was supported by the National Heart, Lung and Blood Institute of the National Institutes of Health (NIH) under award number R01HL122154. At the time this work was conducted, Zainab Farzal was supported by the NIH National Research Service Award Institutional Training Grant 5T32DC005360 to the University of North Carolina. We acknowledge Drs. Landon Holbrook, Kirby Zeman, Jihong Wu, and Chris Jadelis for their expertise on gamma scintigraphy.

Table 1. Subject Demographics

| Subject | Sex | Age | Race | BMI (kg/m2) | Resting Minute Volume (L/Min) | CRS Laterality |
|---|---|---|---|---|---|---|
| **Subject 1** | M | 70 | Caucasian | 24.8 | 9.23 | Bilateral |
| **Subject 2** | F | 24 | Caucasian | 32.6 | 11.81 | Bilateral |
| **Subject 3** | M | 41 | Caucasian | 25.3 | 12.17 | Bilateral |

Table 2. Nebulizer vs. Spray Particle Deposition

|  | NEBULIZER | | | | | SPRAY | | | | |
|---|---|---|---|---|---|---|---|---|---|---|
|  | Particle Size at MSDF (microns) | Fractional Deposition at MSDF (%) | Fractional Deposition at *Spray* MSDF (%) | 50% MSDF Particle Size Range (microns) | Penetrated particles at MSDF particle size (%) | Particle Size at MSDF (microns) | Fractional Deposition at MSDF (%) | Fractional Deposition at *Nebulizer* MSDF (%) | 50% MSDF Particle Size Range (microns) | Penetrated particles at MSDF particle size (%) |
| **Subject 1** | | | | | | | | | | |
| LHS | 13 | 14.6% | 14.0% | 10 to 18 | 96.5% | 12 | 8.0% | 8.2% | 10 to 16 | 20.7% |
| RHS | 21 | 3.7% | 0.5% | 9 to 22 | 55.6% | 4 | 0.6% | 2.1% | 4 to 5, 9 to 12 | 99.8% |
| **Subject 2** | | | | | | | | | | |
| LHS | 11 | 23.1% | 21.0% | 9 to 14 | 92.2% | 10 | 35.3% | 60.5% | 5 to 11 | 94.8% |
| RHS | 15 | 14.6% | 9.8% | 7 to 20 | 82.1% | 7 | 1.3% | 14.7% | 5 to 10 | 94.2% |
| **Subject 3** | | | | | | | | | | |
| LHS | 12 | 7.3% | 7.3% | 9 to 16 | 96.7% | 12 | 6.0% | 6.0% | 10 to 18 | 24.7% |
| RHS | 14 | 14.4% | 14.4% | 11 to 17 | 76.9% | 13 | 8.4% | 9.9% | 11 to 15 | 76.9% |

Table 3. Commercial Nebulizers Available on the Market

| Product | Particle Size (μm) | # of Nasal Applicator Prongs | Compressor Dimensions | Cost | Device Image | |
|---|---|---|---|---|---|---|
| Nasoneb Sinus Therapy System: Model 7070 | 21 | 1 | W 6" x H 7" x D 5" | $120 | 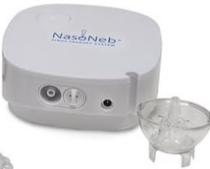 | Source website: http://nasoneb.com/index.php?page=to_use |
| Nasoneb Sinus Therapy System: Nasoneb II, Model 5070 | Not available online | 2 | Not available online | Not available online | 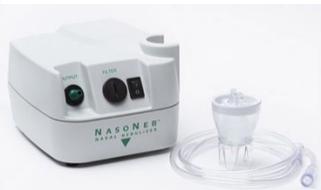 | Source website: http://nasoneb.com/index.php?page=to_use |
| PARI SinuStar | 3 | 2 | 5.75" x 7.5" x 5.5" | $220 | 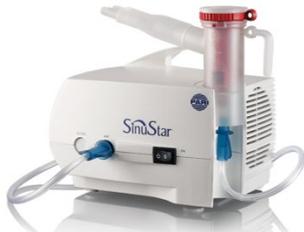 | Source website: https://justnebulizers.com/sinustar-nasal-delivery-system.html |
| PARI SINUS Pulsating Aerosol Compressor System | 3.2 | 1 | W 7.6" x H 5.7" x D 5.9" | $250 | 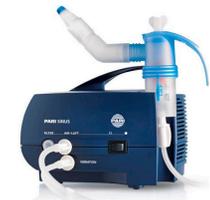 | Source website: http://nebology.com/pari-sinus-pulsating-aerosol-compressor-system.html |

**Figure Legends**

**Figure 1.** Sinonasal 3-D Reconstructions and Mesh. A) Axial View. B) Coronal View. C) Sagittal View. Box denotes OMC region (green) with maxillary sinus removed for visualization. D. Tetrahedral mesh with three prism wide edges at nostril

**Figure 2.** Validation experiment set-up. A) The nebulizer was positioned in a lead-lined box and connected to tubing with an inlet along the 3-D printed model's right nostril surface and held in place with putty. A filter (green) represents the outlet. B, C, and D) In order to assess particle deposition to various regions of the sinonasal cavity, a grid method was used shown.

**Figure 3.** Determination of Particles Meeting 50% Maximal Sinus and OMC Deposition Fraction (MSDF) criteria. After the MSDF and the particle size achieving MSDF were determined, all particles meeting half the MSDF or greater were included and reported as the size range meeting 50% MSDF. This example shows nebulizer particle deposition in subject 1 left hand side (LHS).

**Figure 4.** Determination of Common Nebulizer Particle Size Range Across Subjects. Sinus and OMC deposition is shown for all subjects. The solid color lines represent the deposition fractions for each subject across particle sizes. The matched color dotted lines represent particle deposition fraction at which 50% maximal sinus and OMC deposition fraction (MSDF) was achieved for the corresponding subject. A common particle size range of 11-14 micron particles achieving at least 50% MSDF across all subjects (range represented within the vertical black lines).

**Figure 5.** Nebulizer vs. Spray Particle Maximal Sinus and OMC Deposition Fraction (MSDF). All cases had greater nebulized particle MSDF, except for subject 2 left hand side (LHS) which exhibited greater spray particle deposition.

**Figure 6.** Deposition Fractions for All Sinuses and OMC Across Subjects

**Figure 7.** Post-Nasal Penetrance and Total Deposition Fractions. Post-nasal refers to deposition beyond the anterior nasal cavity region, beyond the internal nasal valve. Penetration fraction = fraction of particles depositing beyond the anterior nasal cavity.

**Figure 8**. Validation of CFD Nebulized Particle Model. Left: Visualization of CFD nebulized particle deposition (blue = deposited particles). Right: Gamma scintigraphy-based nebulized particle deposition (bright signal = deposited particles). "Anterior nose" region is excluded in panels B and C to remove the high gamma scintigraphy anterior nose signal for better visualization of deposition in the regions of interest more posteriorly.

**Figure 9.** Comparison of CFD and *in vitro* deposition patterns for validation. Coronal (A and B), sagittal (C and D), and axial (E and F) compartmental views are shown. The "anterior nose" region was excluded. Compartment labels correspond to consecutive sections of the models in each view as shown in Figure 2.

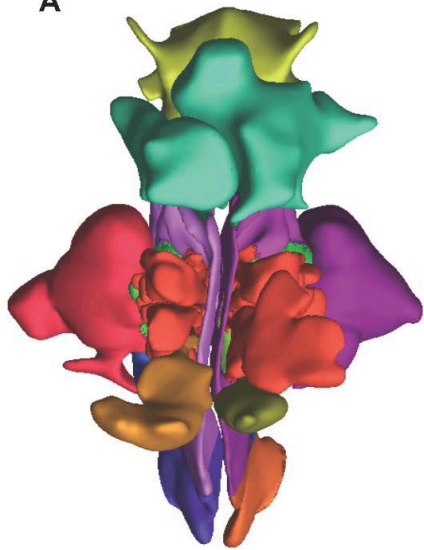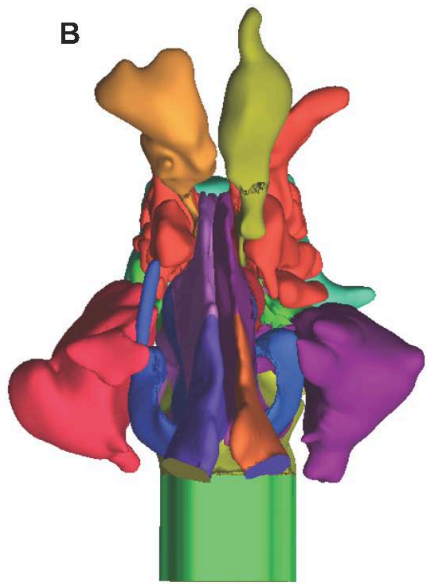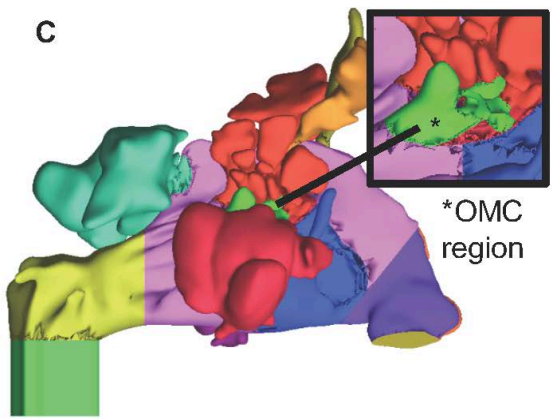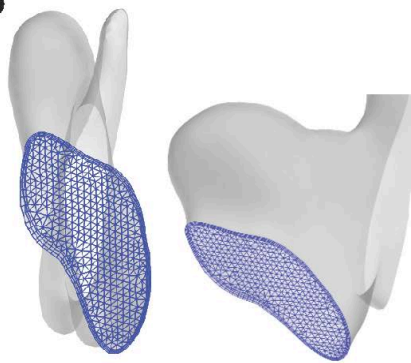

(a) Validation experiment set-up

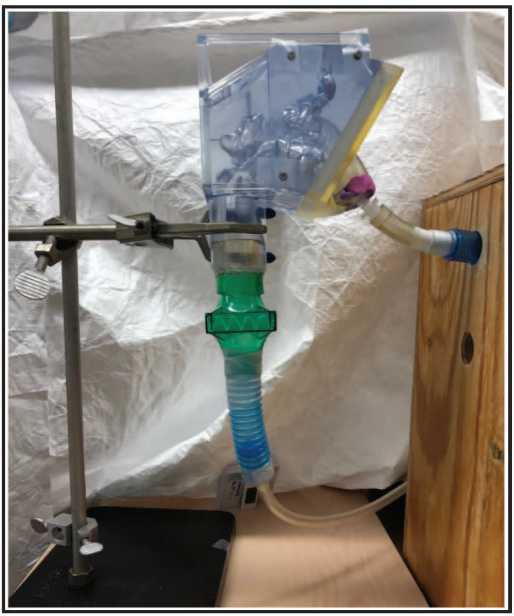
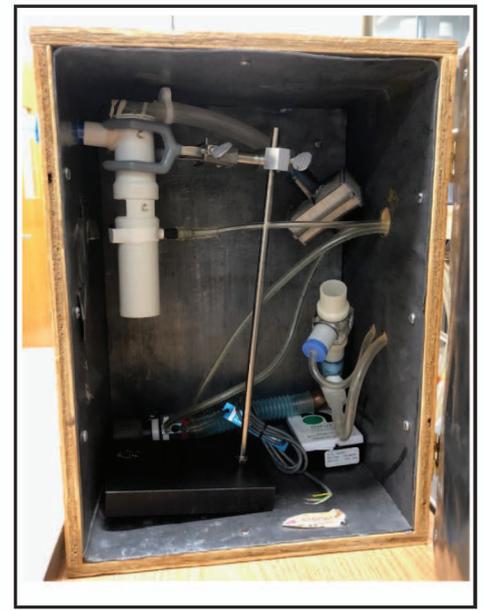

(b) Coronal segments (A1 - A10)

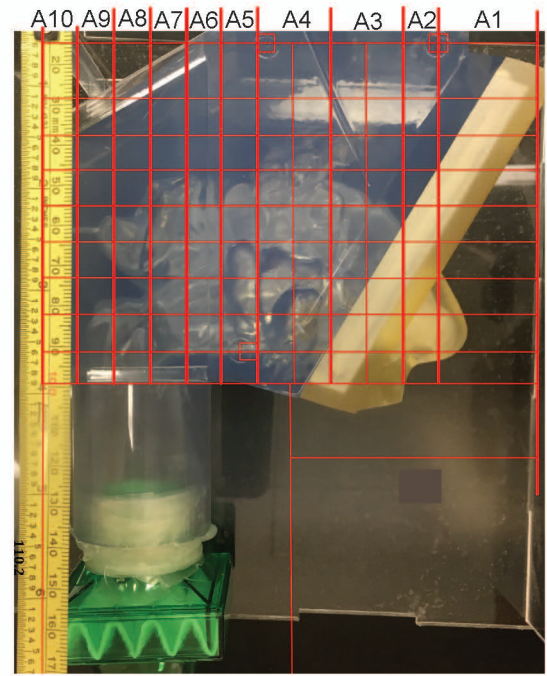

(c) Sagittal segments (B1 - B9)

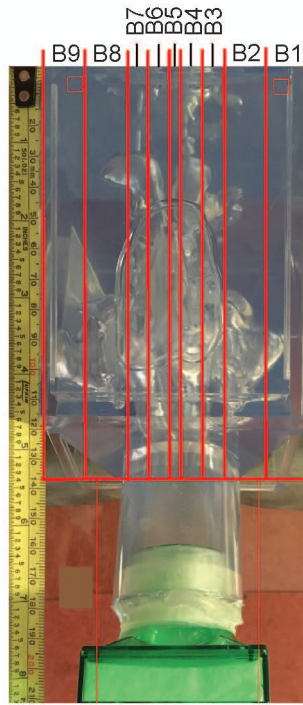

(d) Axial segments (C1 - C12)

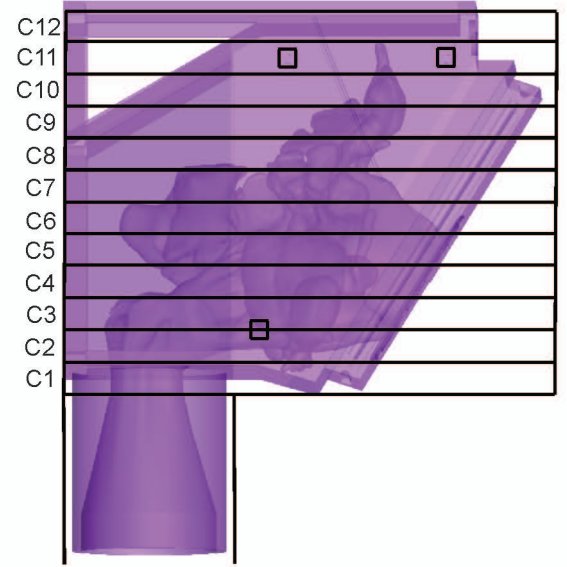

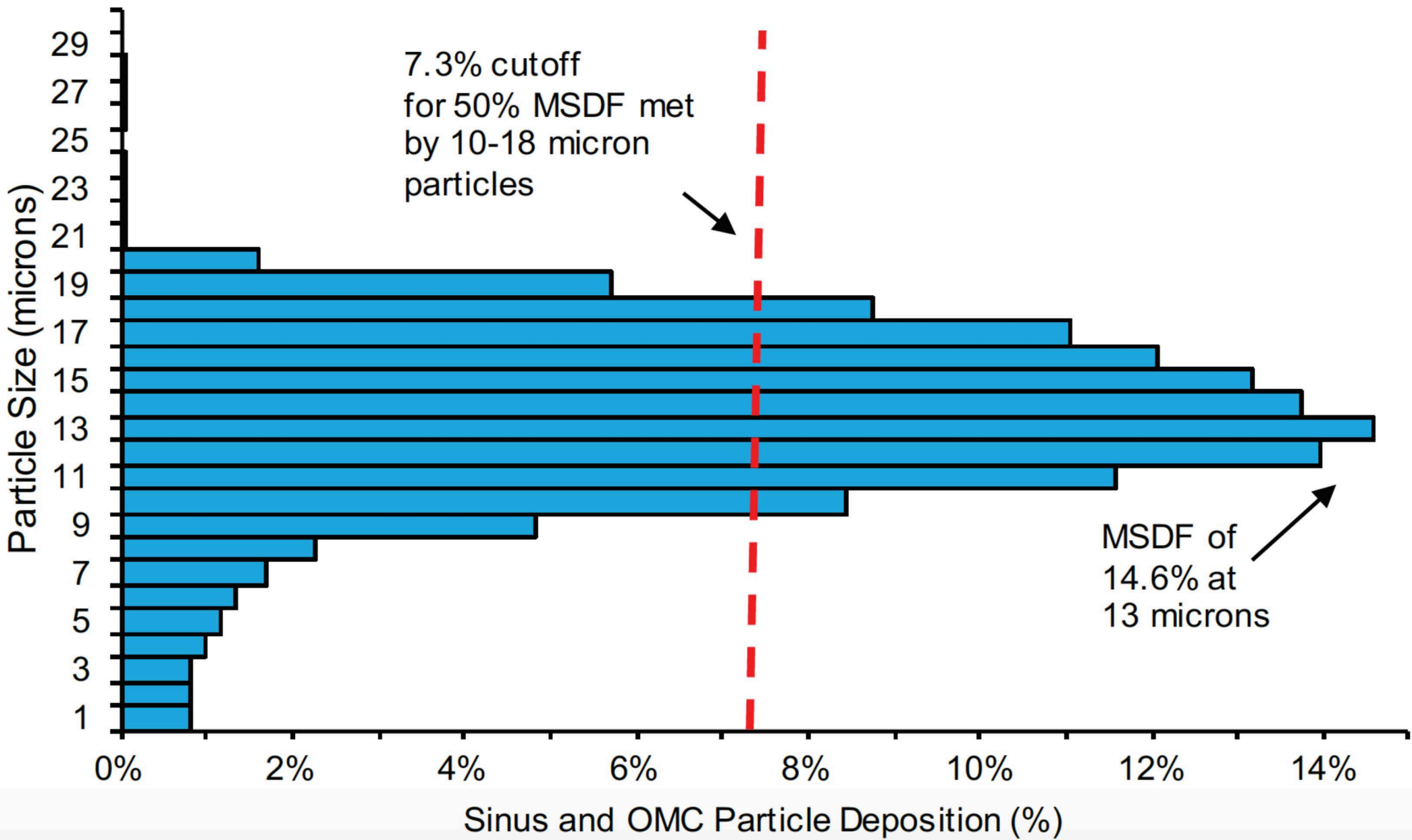

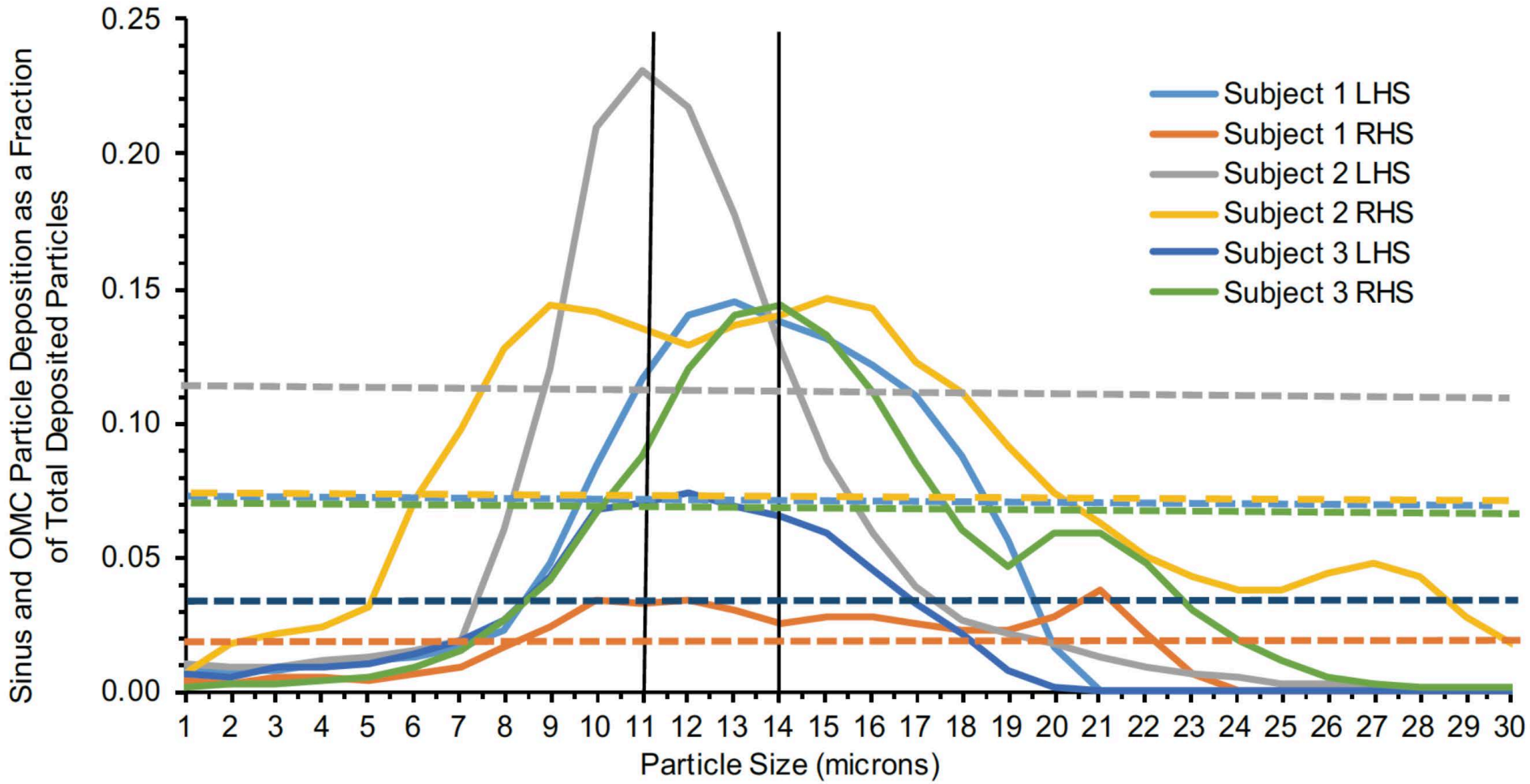

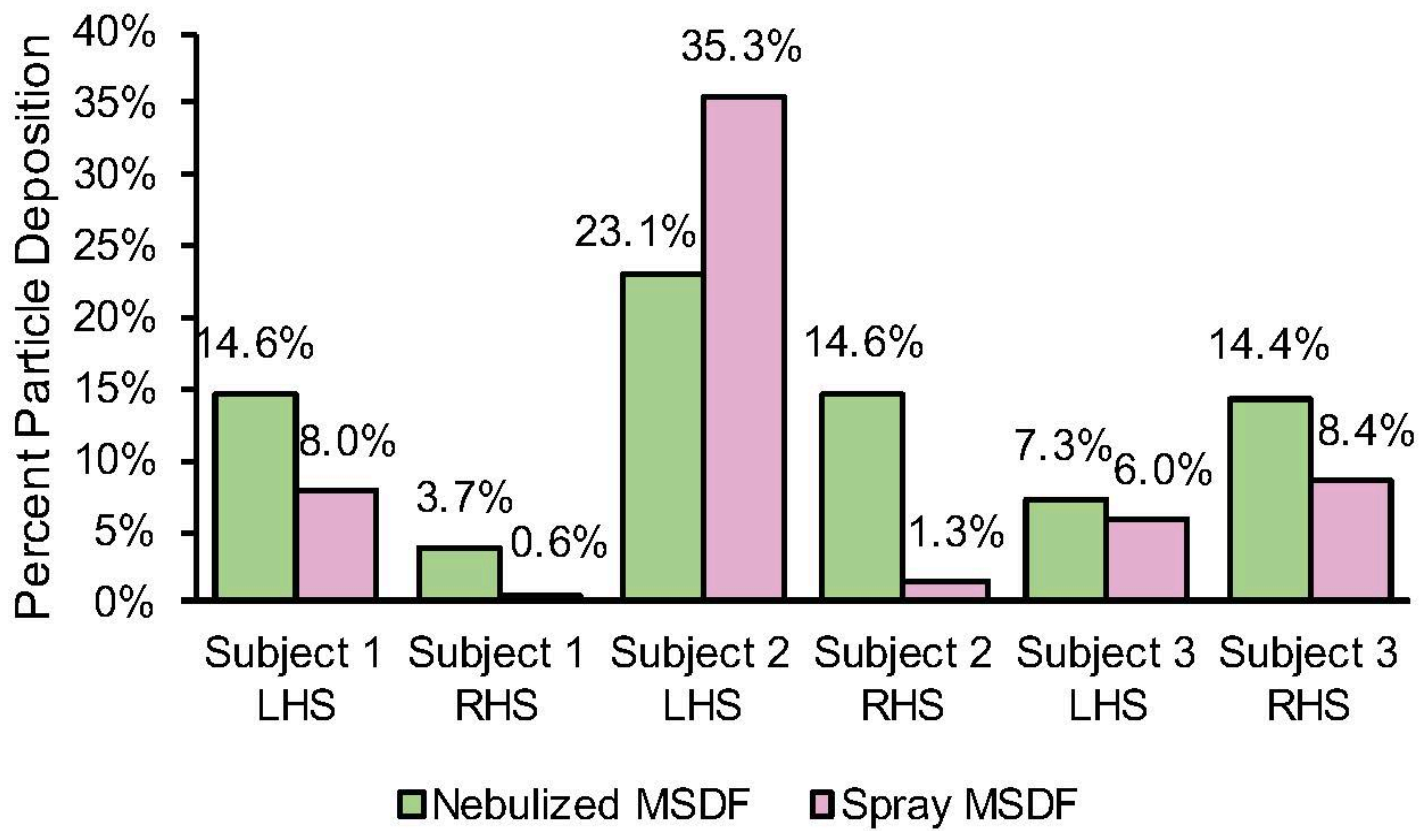

**Subject 1 LHS**

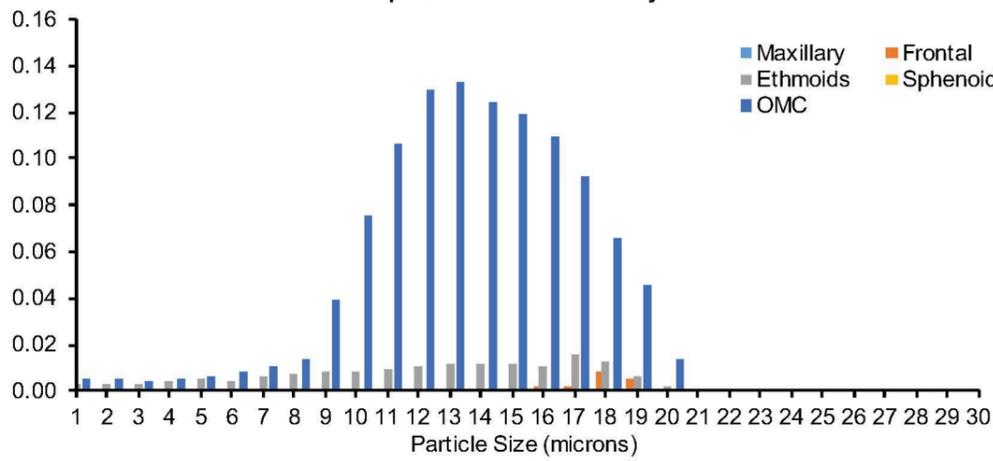
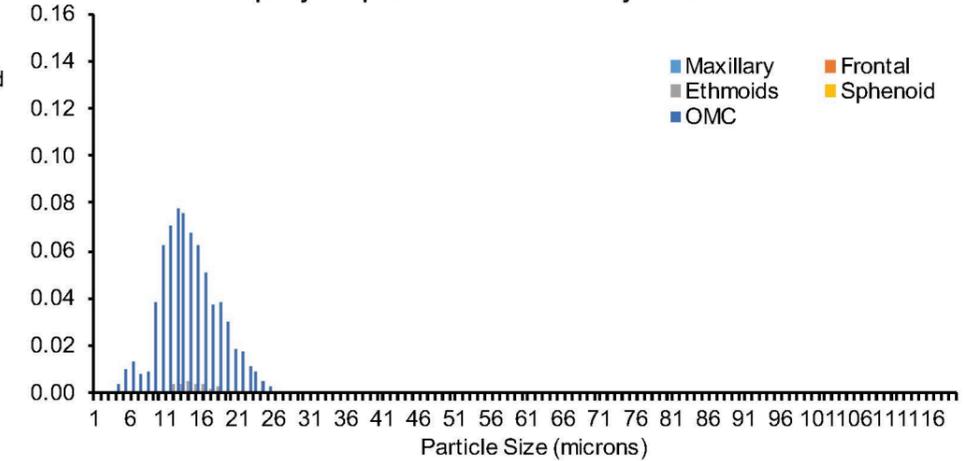

**Subject 1 RHS**

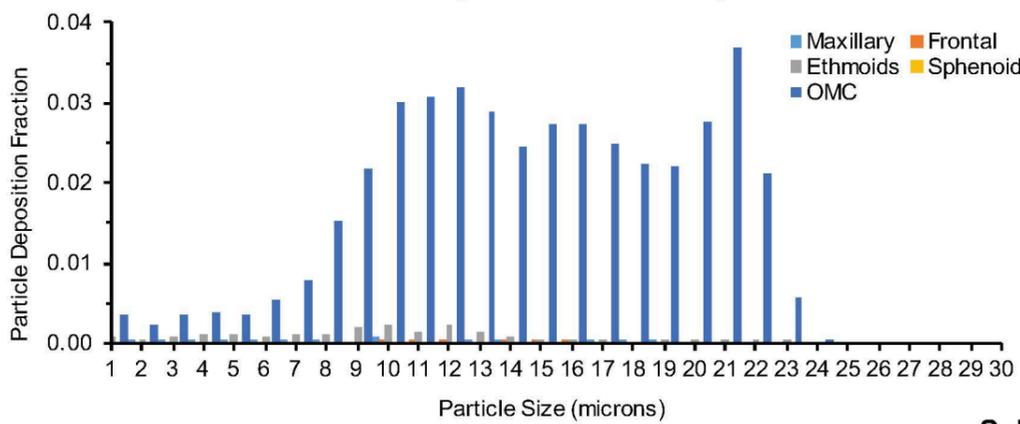
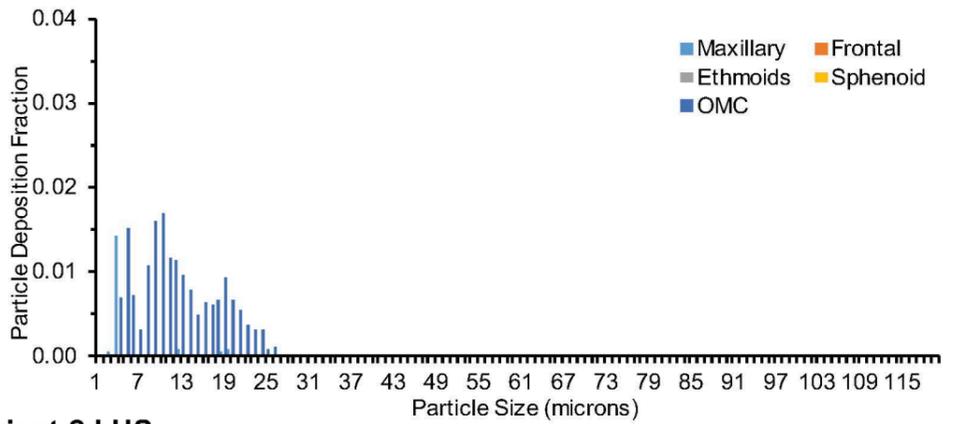

**Subject 2 LHS**

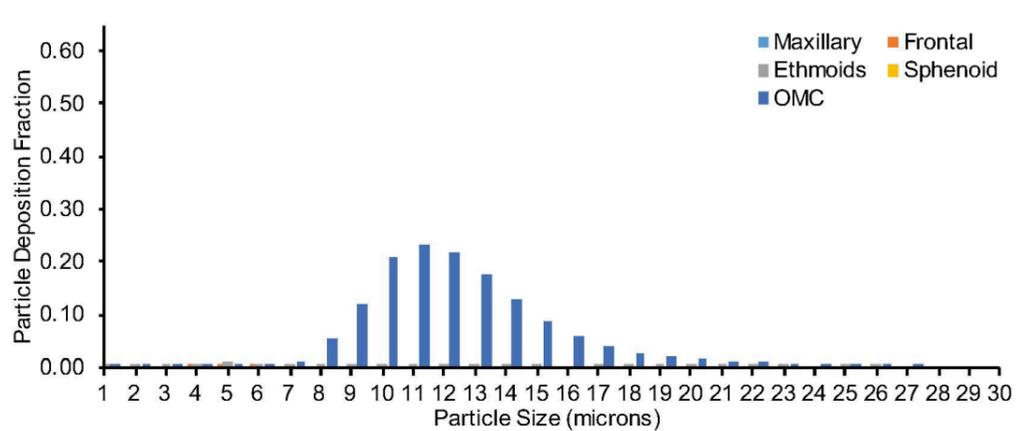
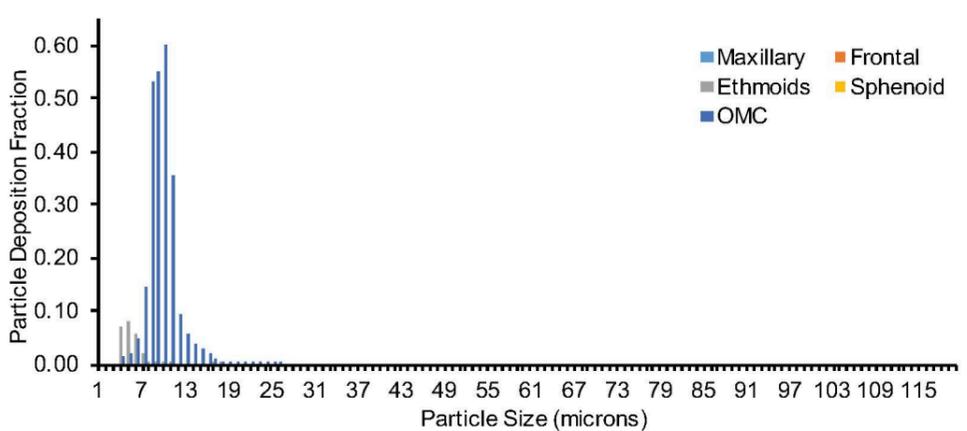

**Subject 2 RHS**

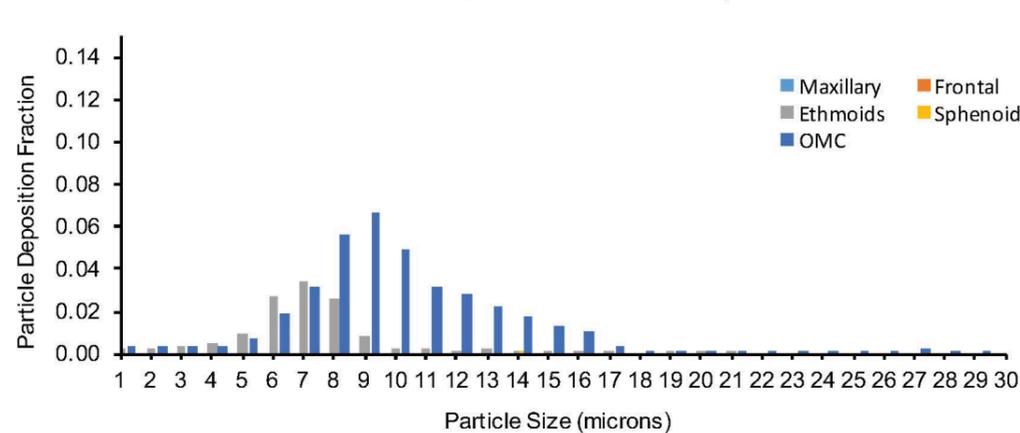
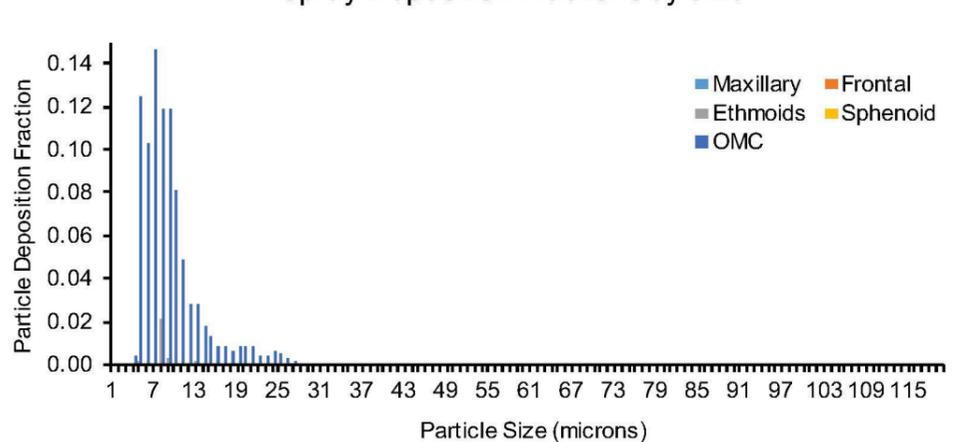

**Subject 3 LHS**

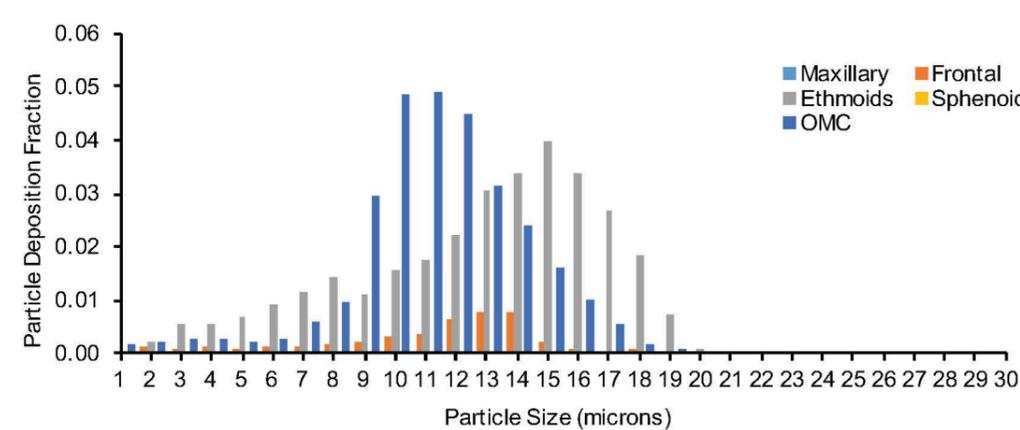
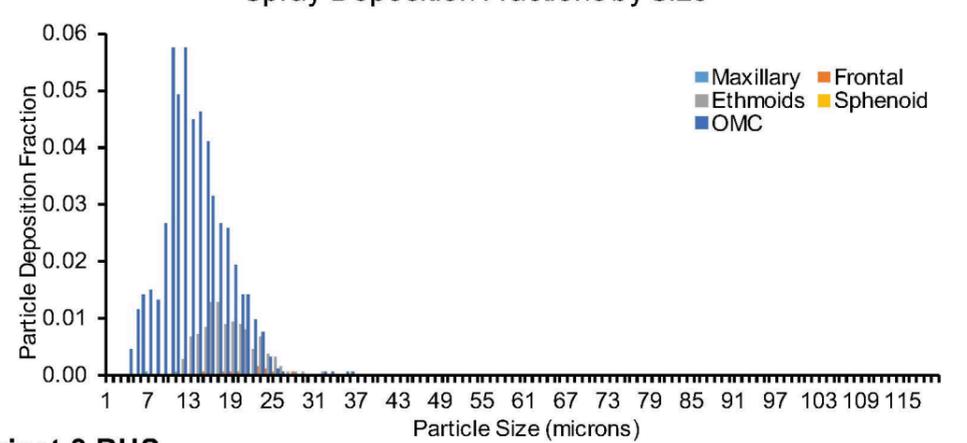

**Subject 3 RHS**

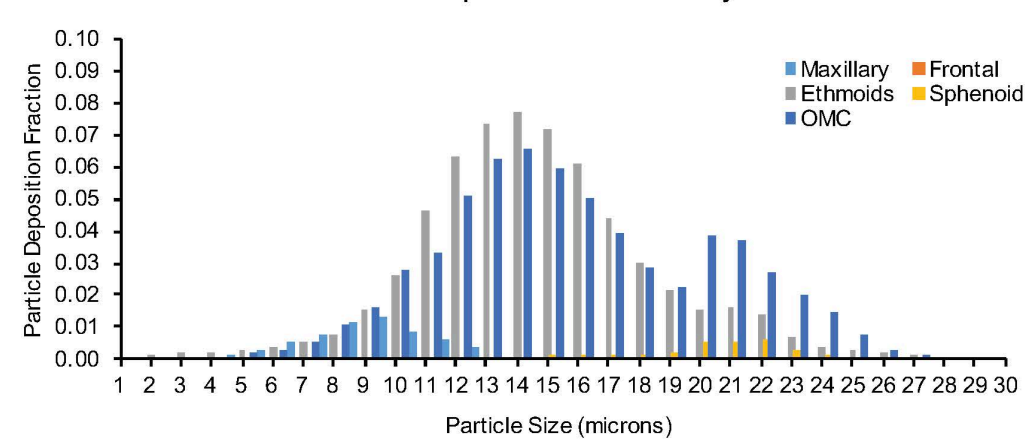
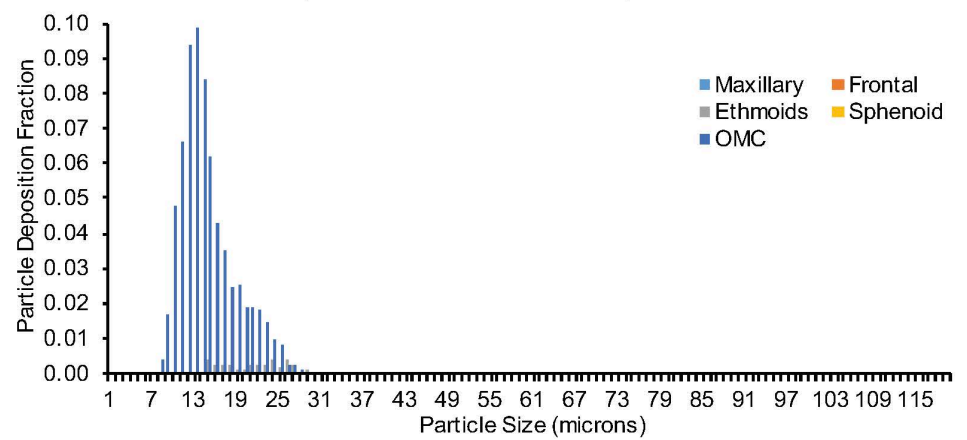

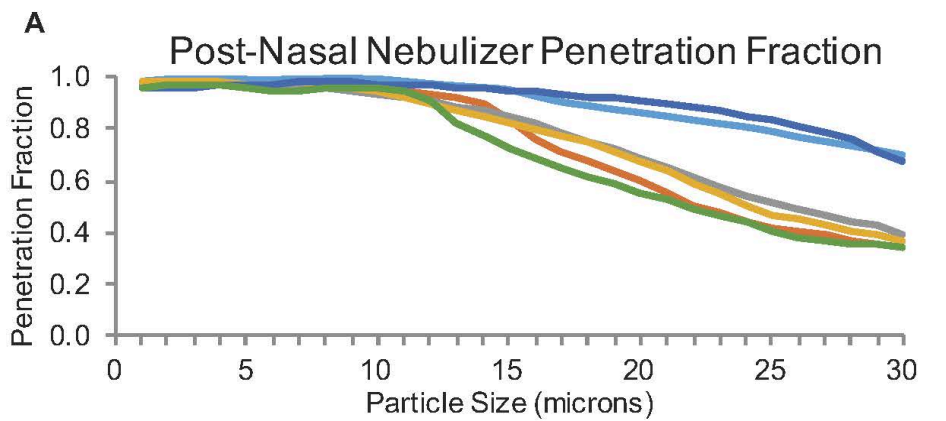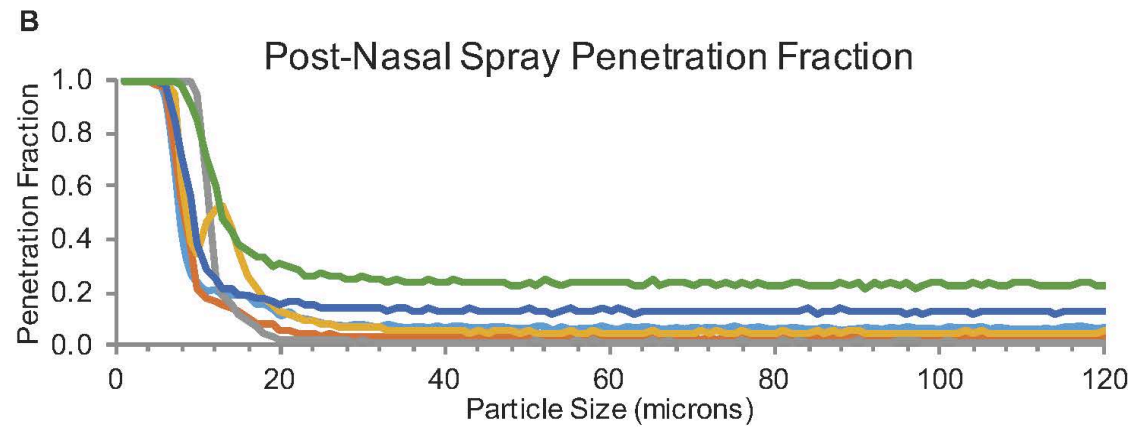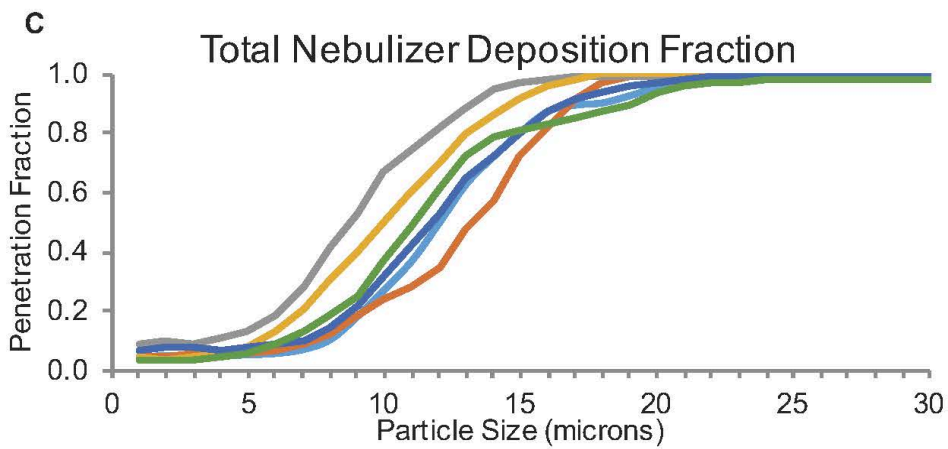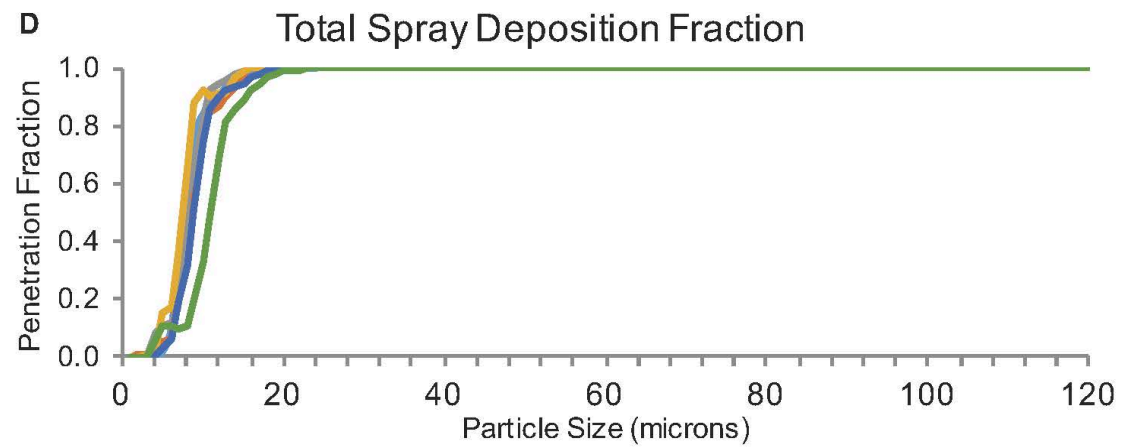

| Simulated through CFD | *In vitro* recordings |

(a) Side view

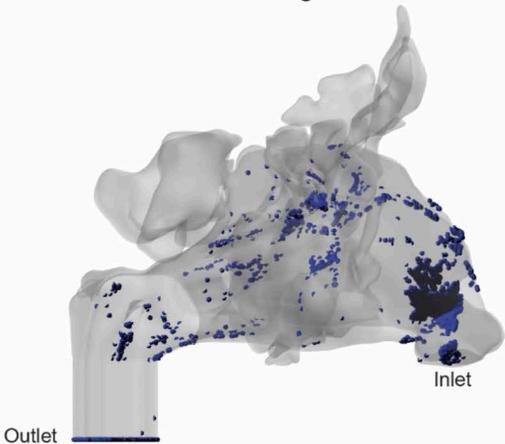 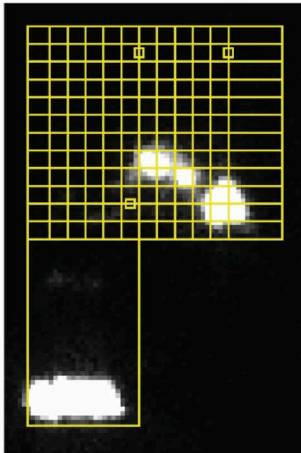

(b) Side view (without anterior nose data)

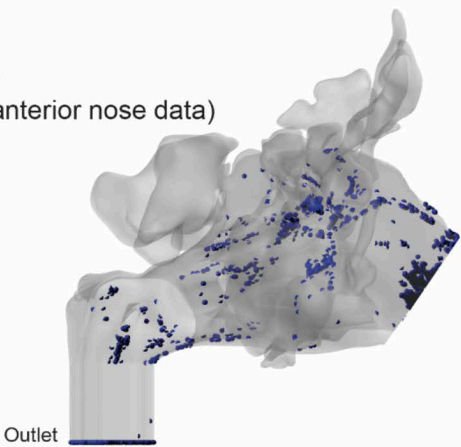 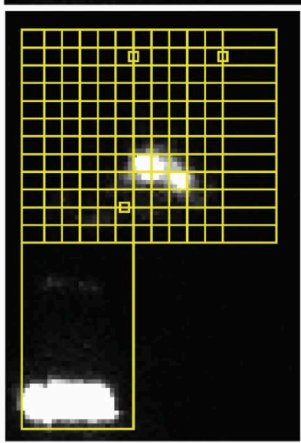

(c) Front view (without anterior nose data)

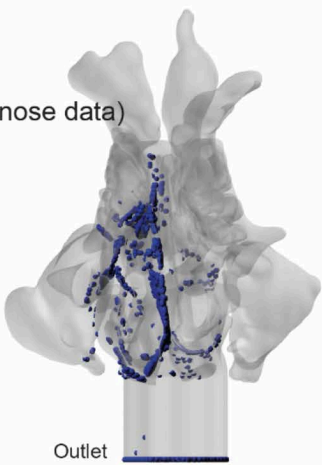 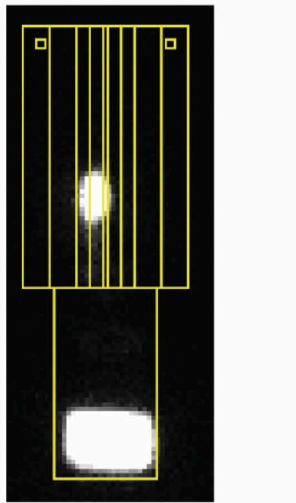

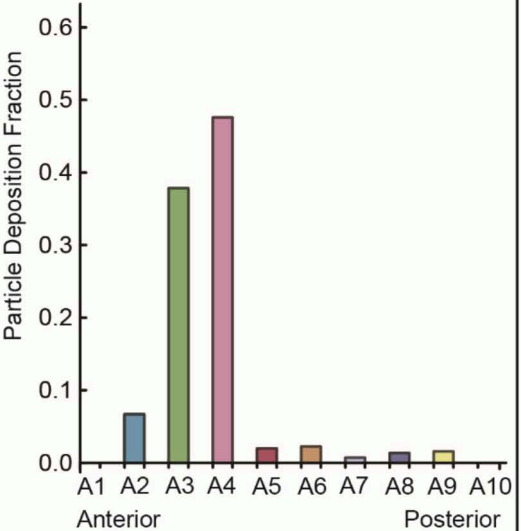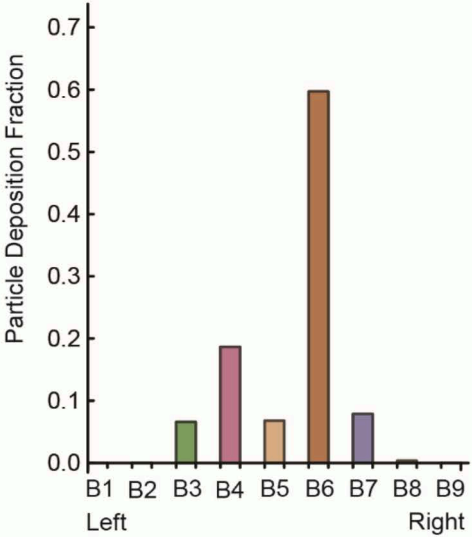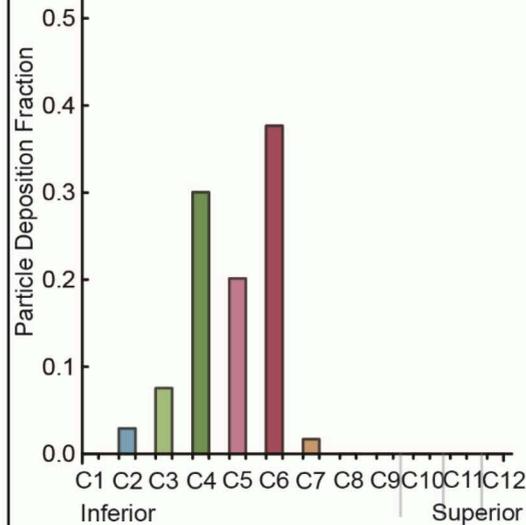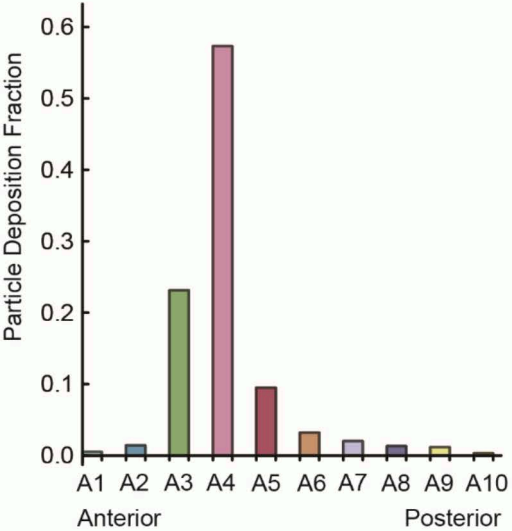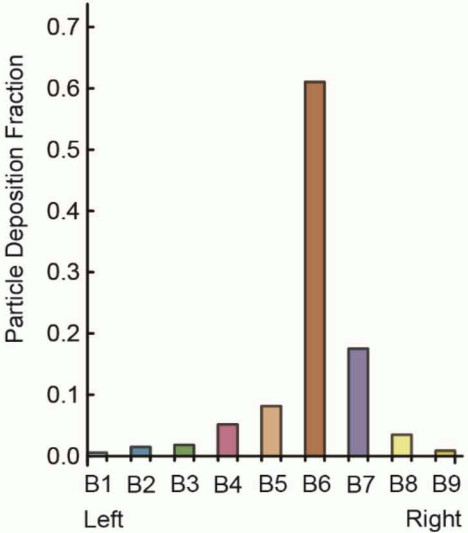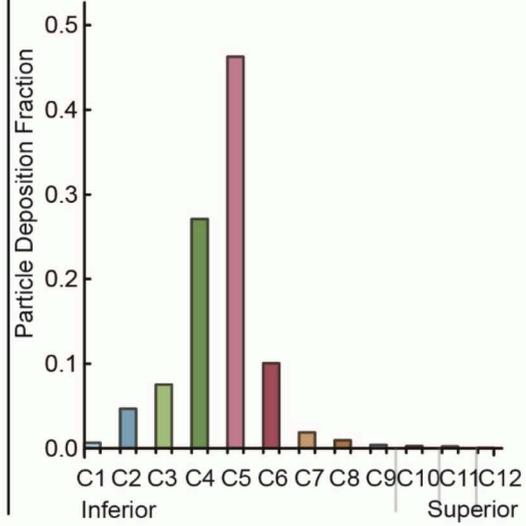